\documentclass[
twocolumn,
superscriptaddress,
 amsmath,
 amssymb,
 aps,
 prl
]{revtex4-1}

\usepackage{graphicx}
\usepackage{subfigure}
\usepackage{dcolumn}
\usepackage{bm}
\usepackage{textcomp}
\usepackage{psfrag}

\begin{document}

\title{Beyond Tanner's Law: Crossover between Spreading Regimes \\ of a Viscous Droplet on an Identical Film}
\author{Sara L. Cormier}
\author{Joshua D. McGraw}
\affiliation{Department of Physics \& Astronomy and the Brockhouse Institute for Materials Research, \\ McMaster University, Hamilton, ON, Canada}
\author{Thomas Salez}
\author{Elie Rapha\"el}
\affiliation{Laboratoire de Physico-Chimie Th\'eorique, UMR CNRS Gulliver 7083, ESPCI, Paris, France}
\author{Kari Dalnoki-Veress}\email{dalnoki@mcmaster.ca}
\affiliation{Department of Physics \& Astronomy and the Brockhouse Institute for Materials Research, \\ McMaster University, Hamilton, ON, Canada}

\date{\today}
\begin{abstract}
We present results on the leveling of polymer microdroplets on thin films prepared from the same material. In particular, we explore the crossover from a droplet spreading on an infinitesimally thin film (Tanner's law regime) to that of a droplet leveling on a film thicker than the droplet itself. In both regimes, the droplet's excess surface area decreases towards the equilibrium configuration of a flat liquid film, but with a different power law in time. Additionally, the characteristic leveling time depends on molecular properties, the size of the droplet, and the thickness of the underlying film. Flow within the film makes this system fundamentally different from a droplet spreading on a solid surface. We thus develop a theoretical model based on thin film hydrodynamics that quantitatively describes the observed crossover between the two leveling regimes.

\end{abstract}

\maketitle


If a liquid droplet is deposited onto a smooth solid surface with a relatively high surface energy, the droplet will spread. Several decades ago, it was shown that the height of the droplet decreases in time as $d_0 \sim t^{-1/5}$ if surface tension is the dominant driving force~\cite{Huh1971,hoffman1975, voinov1976, Tanner1979, Gennes1985, deGennes03Book, blake2006, Croll2009, Courbin2009, bonn2009}. Interestingly, this so-called Tanner's law is valid for any simple liquid that wets a surface. The reason for this universality is that a precursor film emerges ahead of the bulk of the droplet~\cite{Ausserre1986, Brenner1993, cazabat1997, Heine2003, milchev2002, kavehpour2003, bonn2009, Gennes1985, Hoang2011} which consumes the excess surface energy of the solid. The spreading of the droplets that follows the formation of the precursor film is driven by the excess surface area of the nonflat liquid; in essence the droplet is spreading on a thin film of itself. However, as we show here, the power law changes if a droplet spreads on a thicker film that is already present. Specifically, a range of powers can be observed, depending on the precise thickness of the underlying film in comparison with the size of the droplet, with Tanner's law being a limiting case. 

\begin{figure}
\includegraphics[width=1\columnwidth]{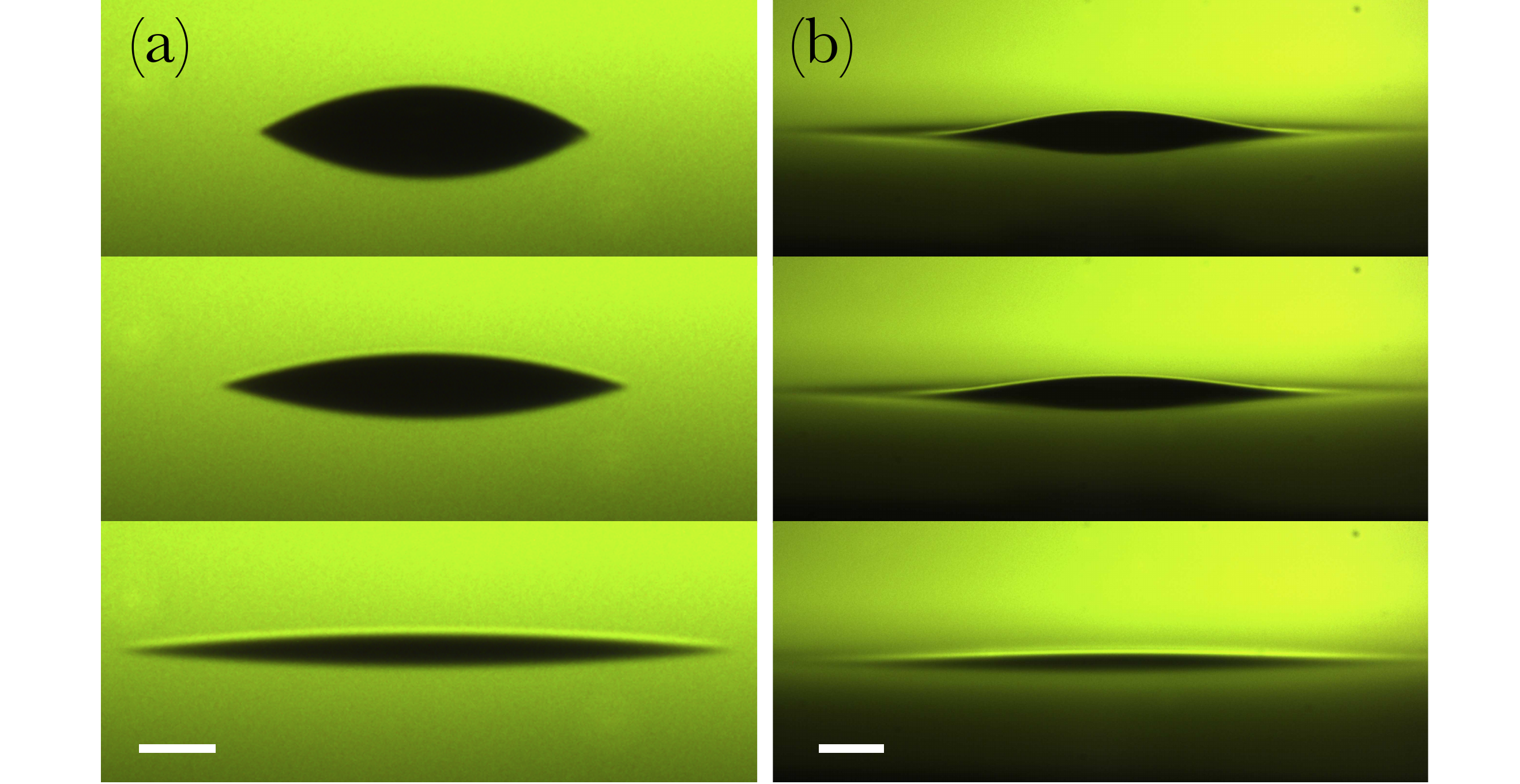}
\caption{Optical microscopy images of silicone oil droplets and their reflection, spreading on a thin film of the same oil. (a)~Droplet spreading on a film with thickness $e \sim 0.2$~\textmu m. (b)~Droplet spreading on a film with thickness $e \sim 22$~\textmu m. The times between the first and last images are $\sim 45$ and $\sim 1$~min; and the scale bars are 100 and 250~\textmu m, respectively, in (a) and (b).
\label{pdms} 
}
\end{figure}

A simple example of a process in which the underlying precursor thickness changes is that of spraying liquid (\emph{e.g.} paint) onto a wetting surface~\cite{jung2012}. The first drops of sprayed paint spread according to Tanner's law~\cite{Tanner1979}, much like the sequence shown in Fig.~\ref{pdms}(a). After the paint droplets have coalesced on the solid surface, a thin film of wet paint will have formed. Due to surface tension, new droplets that are deposited will spread on the thin paint film. The layer of paint deposited in the spray painting process will eventually become thicker than the precursor film (typically 0.1 to 10\ nm thick~\cite{bonn2009}) in Tanner's law spreading. Thus, in addition to the flow within the droplet, we must also consider the flow within the film; this is the situation depicted in Fig.~\ref{pdms}(b). Comparison of the length and time scales in Fig.~\ref{pdms} clearly demonstrates that for every subsequent layer of wet paint that is deposited, the characteristic leveling time of the droplet will change. This time scale will depend on the height of the underlying film and the droplet size.

Small liquid droplets atop identical thin films have been studied computationally~\cite{milchev2002, Pierce2009, Heine2003}. Molecular dynamics simulations have investigated polymer droplets spreading on glassy and liquid polymer substrates of varying molecular weights~\cite{Pierce2009}. It was found that if the polymer droplet and the liquid polymer film are of a different molecular weight, the droplet simultaneously spreads and interpenetrates the polymer film. In addition, these molecular dynamics simulations clearly demonstrated a difference between droplets spreading on a glassy and liquid substrate. In other studies, various spreading laws were found for liquid droplets spreading on an imiscible liquid~\cite{Fraaije1989, Bacri1996, Brochard1996}. Here, we will see that the spreading law can vary even for a droplet spreading on an identical film, the key parameter being the aspect ratio between the drop height and the film height $d_0/e$.

It is the difference between spreading on a thick versus a thin film that is the focus of this Letter. We develop an understanding of the crossover from Tanner's law spreading to the case of a droplet leveling on a film of the same liquid that is relatively thick. Specifically, we investigate how the dynamics of droplet spreading depends on the aspect ratio. The physics is both general and ubiquitous, from the spreading of a droplet of rain on a wet road, to the leveling of a droplet of paint. 

The images in Fig.~\ref{pdms} show the leveling of 100\,000 cs silicone oil droplets (polydimetylsiloxane, Petrarch Systems Inc., USA) with initial radius of curvature $r_c\sim1$~mm. Underlying films were made by spin casting the oil directly onto Si substrates. Droplets of the same oil were deposited onto the film and images were obtained using a simple contact angle measurement set-up. Since the silicone oil is liquid at room temperature, it is difficult to control precisely the initial height profile of droplets on films prepared in this way. 

\begin{figure}
\centering
\includegraphics[width=1\columnwidth]{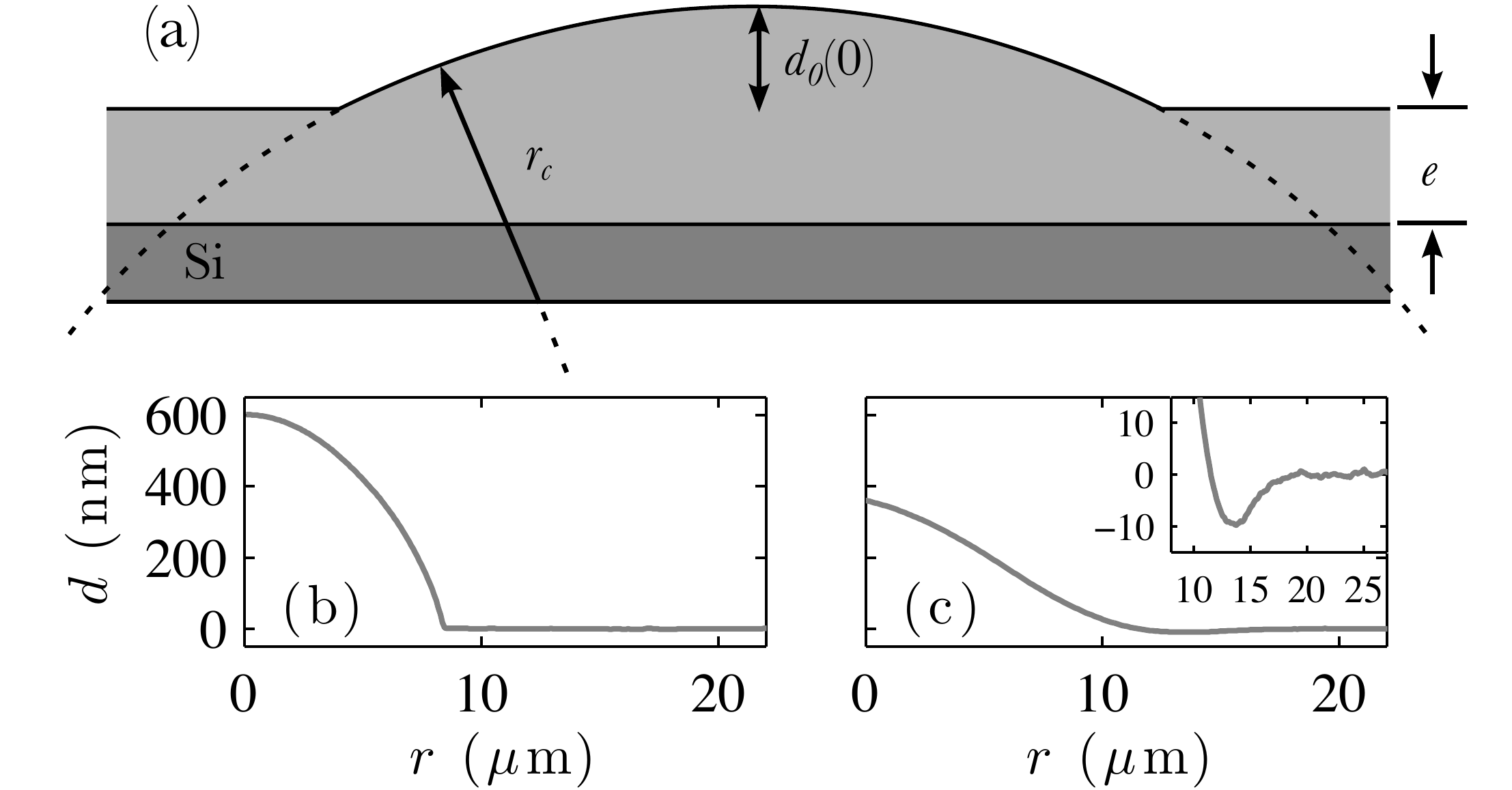}
\caption{(a) Initial geometry: a spherical cap with radius of curvature $r_\textrm{c}$ and height $d_0(0)$ atop a thin film of identical material with height $e$ supported on a Si substrate. (b) AFM line profile $d(r,0)$ of a PS droplet on a thin PS film ($e=201$~nm) before heating ($t=0$) as a function of the distance $r$ to the center of the droplet. (c) AFM profile $d(r,t)$ for the same droplet after heating at 180$^\circ$C for $t=20$~min. The inset shows the same data as in (c) with $d$ and $r$ ranges chosen to emphasize the  dip in the film surface at the periphery of the droplet. }
\label{drawing}
\end{figure}

An ideal initial sample geometry is that of a spherical cap atop a thin film of an identical liquid shown schematically in Fig.~\ref{drawing}(a). 
Samples with such a geometry were prepared using polystyrene (PS), which has a glass transition $T_g \sim 100^\circ\textrm{C}$. The weight averaged molecular weight of the PS was $M_w=118$~kg/mol with a polydispersity index of 1.05 (Polymer Source Inc., Canada). 
PS dissolved in toluene was spin cast onto thin sheets of freshly cleaved mica (Ted Pella Inc., USA). The resulting films were $ \sim 100$~nm thick. Thin films of the same PS in toluene with various concentrations were spin cast onto $10\times10$~mm$^2$ Si substrates. The films on the Si substrates ranged in thickness from $e \sim 50$~nm to $2000$~nm as measured by ellipsometry (EP3, Accurion, Germany). The PS films on mica were placed in a saturated toluene atmosphere. The PS, which is well below the glass transition at room temperature, absorbs toluene, becomes liquid, and readily dewets from the mica substrates. The resulting droplets ranged in size, with typical $r_c\sim100$~\textmu m and height $d_0(0) \sim 1$ \textmu m. Both the droplets on mica and the thin films on Si were annealed in vacuum ($10^{-5}\textrm{ mbar}$) for 24~h at 150$^\circ$C, which is well above $T_g$. The annealing is much longer than the longest relaxation time for the polymer used, removes residual solvent, and forms PS spherical caps on the mica substrate with contact angle less than $15{^\circ}$ as measured with atomic force microscopy (AFM, Veeco Caliber); see Fig.~\ref{drawing}(b). 

At room temperature, with PS in the glassy state, the entire mica sheet was dipped into clean water (Milli Q, 18.2 M$\Omega$\,cm) and droplets floated atop the surface due to the hydrophobicity of PS. The film of PS on Si was skimmed across the surface of water containing the floating droplets. Droplets were thus picked up by the PS film resulting in the desired sample configuration as shown in Fig.~\ref{drawing}(a). With this sample preparation, we were able to easily control the initial aspect ratio of the system, which had a range of $0.1 \leq d_0(0)/e \leq 100$. Though we focus here on the simplest symmetric system for the droplet and film, polymers of different $M_w$, or different chemical composition can also be prepared. 

To observe droplet leveling, the samples were rapidly heated to 180$^\circ$C, well above $T_g$, on a heating stage (Linkam, UK). All experiments with PS were performed using optical microscopy in reflection (Olympus, USA) and illuminated with a red laser line filter ($\lambda = 632.8$~nm, Newport, USA). Images were obtained for 1 to 17~h,  depending on the initial aspect ratio. Once PS is in the melt state, the Laplace pressure acts to reduce the curvature gradients and the droplet levels (see video in the supplemental material~\cite{Note1}). 

Initially, at the contact line where the spherical cap meets the PS film, there is a high negative radial curvature. This curvature is sharpest prior to heating, as can been seen in the AFM line profile shown in Fig.~\ref{drawing}(b), where the droplet profile $d(r,0)$ is shown as a function of the radial position $r$ from the centre of the droplet. The pressure gradients result in a dip around the circumference of the drop [Figs.~\ref{pdms}(b) and~\ref{drawing}(c)]; similar features appear in the relaxation of stepped surface profiles~\cite{McGraw2011} and were predicted in droplets on films by Tanner~\cite{Tanner1979}. At early times, the curvature gradients are strongest near the dip, and this region dominates the flow. At later times, the curvature at the top of the droplet dominates and the entire droplet levels. In the present Letter, we describe in detail this latter stage of droplet leveling.

Figures~\ref{pdms} and~\ref{drawing} illustrate the role of the underlying film. A droplet with a large height $d_0$ in comparison to the film thickness $e$, spreads much like Tanner's wetting droplet [see Fig.~\ref{pdms}(a)]. In this case the infinitesimal film is like a precursor film. In contrast, a droplet on a film that is thicker spreads in a way that is qualitatively different. From Figs.~\ref{pdms}(b) and~\ref{drawing}(c) it is clear that the liquid surface of the film is perturbed well beyond the
position of the initial contact line. Furthermore, the characteristic leveling times in Fig.~\ref{pdms} are vastly different: though in the first frames shown the droplet heights differ by only a factor of $\sim 3$, the time scales in the sequence vary by a factor of $\sim 50$. 

\begin{figure}
\centering
\includegraphics[width=1\columnwidth]{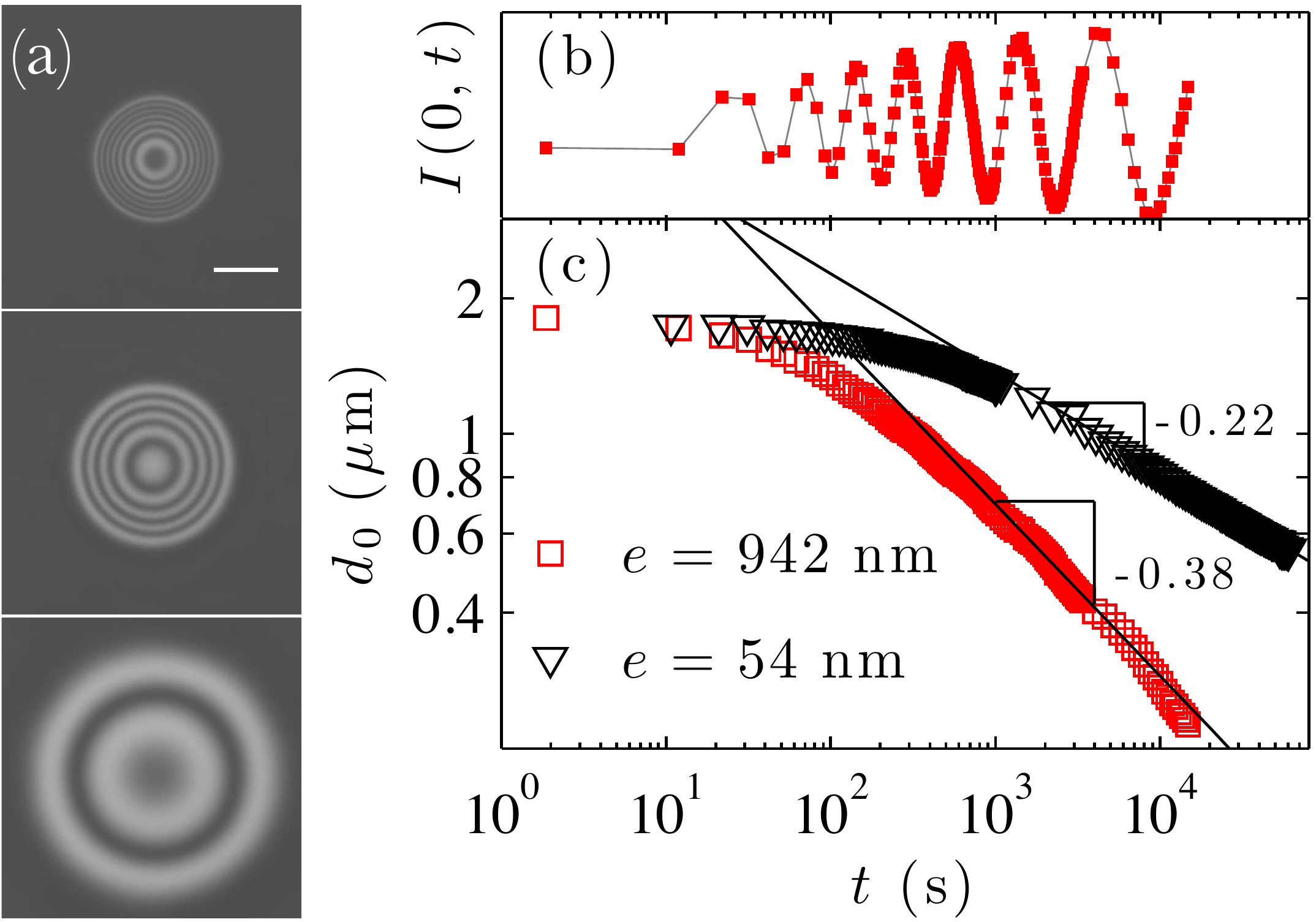}
%
\caption{
\label{newton} 
(a) Time series ($\sim$5~h elapsed time) optical microscopy images showing the Newton rings of a droplet spreading on a film with $e$ = 112 nm (10 \textmu m scale bar). (b) Intensity at the centre of the droplet as a function of time. (c) Temporal power-law decrease in droplet height for two independent droplet experiments. }
\end{figure}

We now turn to a full description of the experiments using PS droplets on PS films. In Fig.~\ref{newton}(a) we show optical microscopy images of a droplet spreading over $\sim$5~h of annealing. Newton rings are observed as a result of interfering monochromatic light reflecting from both the air-PS and PS-Si interfaces. The intensity $I(r,t)$ changes periodically with total height $h(r,t)$ with respect to the substrate, as described by the optical thin film equations~\cite{Heavens1960a}. We monitor the change in the intensity at the centre of the drop with time, as seen in Fig.~\ref{newton}(b), and determine the central height $h(0,t)$~\cite{Note2}. Figure~\ref{newton}(c) shows $d_0(t) =d(0,t)= h(0,t)-e$ for two independent droplet leveling experiments with different aspect ratios, $d_0/e$. Comparing the two experiments, we observe that the long time data can be described by two different power laws. If the droplet height is large in comparison to the film (triangles), we observe a spreading law that approaches Tanner's law: $d_0 \sim t^{-n}$, with $n = 0.22 \approx 1/5$. In the limit $d_0/e \rightarrow \infty$, the thin film is thus identical to the precursor film of a Tanner's law droplet. On the other hand, when $d_0/e$ is small (squares), we observe spreading with a stronger time dependence than Tanner's law, characterized by $d_0 \sim {t^{-n}}$ with $n = 0.38 > 1/5$. The system evolves faster when the film height is larger.

In the following, we derive a theoretical model in order to unify the experimental results in all regimes. First, since the height scale is small compared to typical radial scales, the lubrication approximation~\cite{Oron1997, RevModPhys.81.1131,deGennes03Book} can be used. Second, the droplet leveling is driven by gradients in the Laplace pressure. This pressure is proportional to both the surface tension, $\gamma$, at the air-fluid interface, and the two principal curvatures. The height scales involved in these experiments allow us to neglect the hydrostatic~\cite{deGennes03Book, brochwy1991} and disjoining pressures~\cite{Perez2001, ghosh2010}.
Third, since we consider a highly viscous fluid with viscosity, $\eta$, it is reasonable to use the Stokes equation to connect the local velocity and pressure. Finally, we assume incompressibility of the flow. Therefore, according to the cylindrical geometry of the system studied, the capillary driven thin film equation
\begin{equation}
\partial_t h+\frac{\gamma}{3\eta}\frac{1}{r}\partial_r\left[rh^3\left(\partial^3_rh+\frac{1}{r}\partial^2_rh-\frac{1}{r^2}\partial_rh\right)\right]=0\ ,
\label{lubrication}
\end{equation}
describes the evolution of the total height of the free interface: $h(r,t)=e+d(r,t)$. In addition, we assume that the underlying film of constant thickness $e$ is infinite in lateral extent. Compared to the analysis of~\cite{Stillwagon1990, Stillwagon1988}, the geometry of our system requires us to retain the last two terms in Eq.~(\ref{lubrication}). 

We nondimensionalize the problem through: $D=d/e$, $D_0=d_0/e$, $R=r/e$, $R_\textrm{c}=r_\textrm{c}/e$ and $T=\gamma t/(3\eta e)$, and we assume the existence of self-similar solutions of the second kind of the form~\cite{barenblatt}:
\begin{equation}
\label{ss1}
D(R,T)=D_0(T)F(U)\ , 
\end{equation}
where
\begin{equation}
\label{ss2}
U=\frac{R}{T^{1/4}\left[1+D_0(T)\right]^{3/4}}\ .
\end{equation}
This particular horizontal scaling is analogous to the one proposed in~\cite{Stillwagon1988}, except that we take into account the time dependence of the typical height through $D_0(T)$. Note that this scaling is not valid at early times for the reason given above: the short time dynamics is governed by the radial curvature at the edge of the droplet. In the following, we assume that such a self-similar solution is reached after a certain time depending on the initial geometry. The self-similarity has been experimentally verified using AFM profiles such as the one in Fig.~\ref{drawing}(c). 

The additional volume $V$ with respect to the volume of the underlying film is constant by incompressibility, and equals $V=2\pi e^3\int_0^{\infty}dR\ R\ D(R,T)$.
Using Eqs.~(\ref{ss1}) and~(\ref{ss2}) we get the time evolution of the droplet height
\begin{equation}
\label{model}
\left(\frac{d_0}{e}\right)^2\left(1+\frac{d_0}{e}\right)^3=\frac{\tau}{t}\ .
\end{equation}
The characteristic leveling time, $\tau$, is given by
\begin{equation}
\label{tau}
\tau=\frac{1}{\alpha^2}\ \frac{\eta V^2}{\gamma e^5}\ ,
\end{equation}
where 
\begin{equation}
\alpha=\frac{2\pi}{\sqrt{3}}\int_0^{\infty}dU\ U\ F(U),\ 
\label{alph}
\end{equation}
is a geometrical factor depending only on the dimensionless height $D_0(0)$ and radius of curvature $R_\textrm{c}$ of the initial spherical droplet. 
If we examine the limits of Eq.~(\ref{model}), we find that for $d_0 \gg e$ Tanner's law is recovered~\cite{caveat}: $d_0 \sim{t^{-1/5}}$. In contrast, for $d_0 \ll e$ we get $d_0\sim{t^{-1/2}}$: the model predicts that a droplet of a given size will level faster on a thicker film, in accordance with the simple observations made for the silicone oil droplet sequences in Fig.~\ref{pdms} and the PS data in Fig.~\ref{newton}(c)~\cite{eggers}.

\begin{figure}
\includegraphics[width=1\columnwidth]{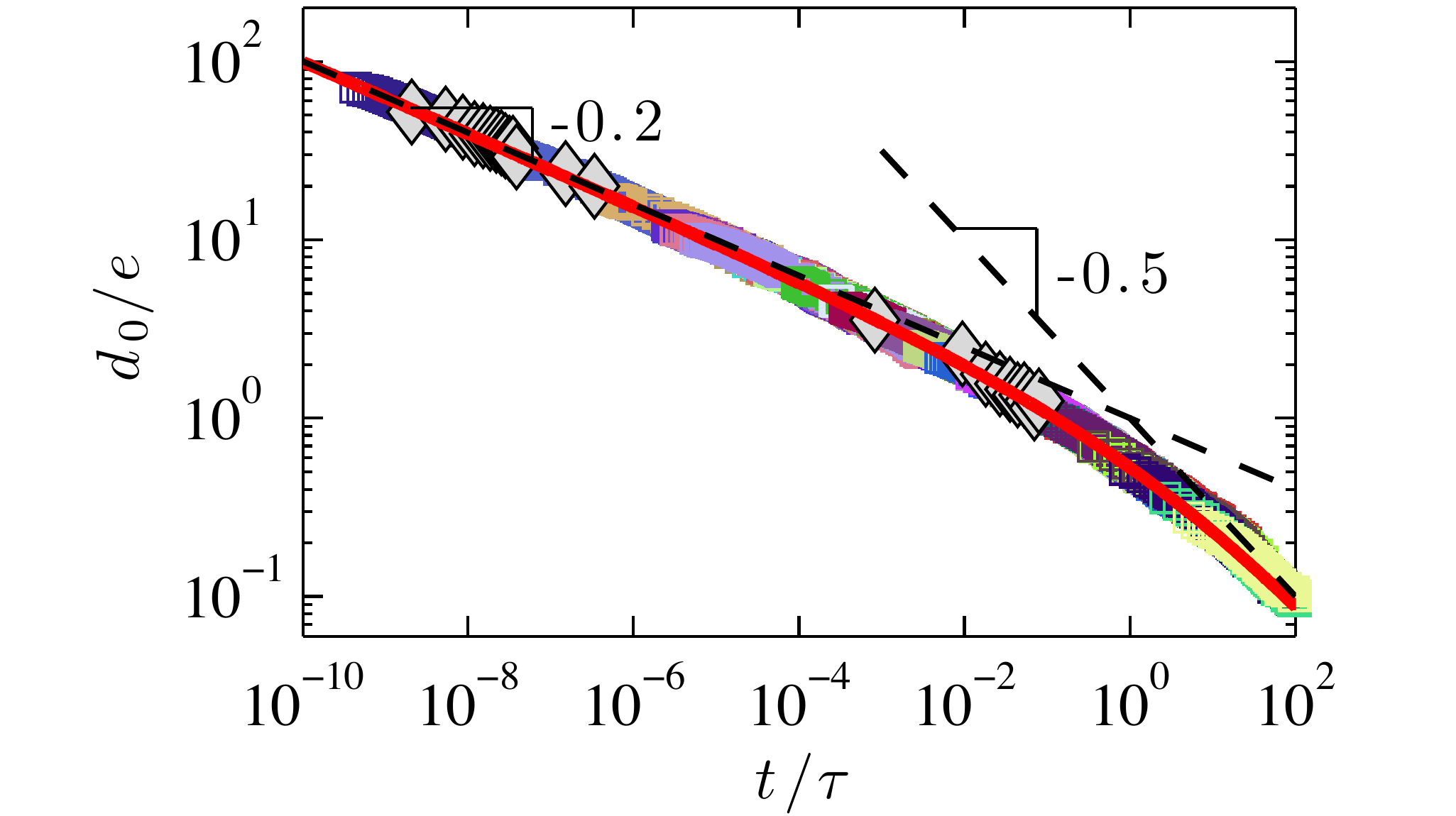}
\caption{\label{master}Plot of the droplet-film aspect ratio as a function of the renormalized time defined by Eq.~(\ref{tau}). The plot contains 35 experiments with PS droplets (squares, various colours) as well as the curve corresponding to the model of Eq.~(\ref{model}) (red solid line). The Tanner limit ($d_0 \gg e$) and thick film limit ($d \ll e$) are shown with dashed lines. Grey diamonds correspond to the data for the silicone oil droplet experiment from Fig.~\ref{pdms}.}
\end{figure}
To test the model prediction of Eq.~(\ref{model}), we performed 35 independent droplet leveling experiments of the type shown in Fig.~\ref{newton} using PS. For each experiment the initial aspect ratio $d_0(0)/e$ and volume  $V$ were different. The evolution of $d_0$ was then compared to Eq.~(\ref{model}). For each sample, we determine $V$ using the optical microscopy images, and $e$ as described above. In addition, we measure $\gamma/\eta = 3.2\ \textrm{\textmu m}/\textrm{s}$ using the stepped film analysis of \cite{McGraw2012}. Therefore, $\alpha$ in Eq.~(\ref{alph}) is the only adjustable parameter in a fit to Eqs.~(\ref{model}) and~(\ref{tau}). We find experimentally that $4 < \alpha < 15$ for the studied samples and that $\alpha$ varies monotonically with the aspect ratio. Using the experimentally determined aspect ratio and radius of curvature, we also calculate $\alpha$ from numerical solutions to the nonlinear partial differential  Eq.~(\ref{lubrication})~\cite{Bertozzi1998,Salez2012a}. The experimentally and numerically determined values agree and both the spread and trend observed are consistent with the theory. The results of all PS droplet leveling experiments in this study are plotted in Fig.~\ref{master} where it can be seen that the data accesses 12 orders of magnitude in scaled time~\cite{eggers}. Moreover, the result clearly shows that the droplet leveling dynamics depends sensitively on the aspect ratio of the system through the flow in the underlying film.

Since the previous model assumes only that the leveling is driven by the Laplace pressure and that the thin film equation applies, the result has to be general and should hold for any thin and nonvolatile viscous fluid. To test this statement, the preliminary silicone oil data from Fig.~\ref{pdms} was added to the master plot in Fig.~\ref{master} (diamonds). In comparison to the experiments with PS droplets, the oil is 2 orders of magnitude less viscous and the characteristic length scales 2 orders of magnitude larger; nevertheless, both the silicone oil and PS data collapse, consistent with the theoretical prediction.

In this Letter, we have presented our work on the leveling of a viscous droplet on a film of the same liquid. It was found that the spreading law is controlled by the aspect ratio of the droplet height to the film height. To describe this spreading dependency, we developed a model based on a Laplace pressure driven viscous flow. The model predicts a crossover in the power law of spreading from the $d_0 \gg e$ Tanner regime, with $d_0\sim{t^{-1/5}}$, to the case of $d_0 \ll e$ for a droplet leveling on a thicker film, with $d_0\sim{t^{-1/2}}$~\cite{caveat}. The model has been verified by experiments carried out using PS and silicone oil droplets, with initial aspect ratios ranging from $d_0(0)/e \sim 0.1$ to 100. The crossover from Tanner's law spreading to droplets leveling on films is a general result that holds for any nonvolatile fluid in the thin film geometry. 

The authors thank NSERC of Canada, the \'Ecole Normale Sup\'{e}rieure of Paris, the Chaire Total-ESPCI, and the Saint Gobain Fellowship for financial support.


\begin{thebibliography}{37}
\bibitem{caveat} Note that our model does not describe the microscopic details at the contact line when $e$ approaches molecular length scales.
\end{thebibliography}
%

\end{document}